# Application-Based Evaluation of an Efficient Spike-Encoder, "Spiketrum"

MHD Anas Alsakkal, Runze Wang, Huajin Tang, and Jayawan Wijekoon, Member, *IEEE*

*Abstract*— **Spike-based encoders represent information as sequences of spikes or pulses transmitted between neurons and have demonstrated strong capabilities in capturing spatiotemporal dynamics while offering energy-efficient solutions for low-power applications. The Spiketrum encoder efficiently compresses input data using spike trains or code sets (for non-spiking applications) and is adaptable to both hardware and software implementations, with lossless signal reconstruction capability. This paper proposes and evaluates Spiketrum hardware under varying spike rates, assessing its classification performance with popular spiking and non-spiking classifiers, as well as the quality of information compression and hardware resource utilization. Extensive benchmarking is performed against state-of-the-art biologically plausible encoders using criteria including classification accuracy, training speed, sparsity, encoded output entropy, and hardware resource utilization and power consumption. Results demonstrate Spiketrum's superiority across most benchmarking criteria, making it a promising choice for various applications with efficient hardware utilization, low power consumption, and high classification accuracy. In addition, this paper proposes an adaptive learning approach to obtain application-specific kernel sets tailored to signals with different spatiotemporal characteristics. Its applicability is demonstrated using electrocardiogram (ECG) signals, extending the Spiketrum framework beyond audio processing. This work also emphasizes the potential of encoders in spike-based processing to improve the efficiency and performance of neural computing systems.**

## I. Introduction

To attain energy-efficient brain-like intelligence computing, a promising approach would be to build dedicated neuromorphic spike-based hardware that closely emulates the brain's processing and biological sensing principles. Many of the existing acoustic spike-based encoding algorithms are either hardware-efficient but do not closely follow the functionality observed in the ear, or they are biologically-plausible but cannot efficiently be realized in hardware. Most of the hardware-efficient spike-based algorithms extract features based on spectrogram representations (time vs frequency plots) of acoustic signals [1-3]. However, such techniques used by the algorithms cannot efficiently extract both temporal and spectral information from harmonic-rich acoustic signals, resulting in significant information loss during the waveform-to-spike transform process. On the other hand, biologically-plausible models mimic most of the relevant physiological functions found in the cochlea, such as the function of the inner hair cells, outer hair cells, or the ribbon synaptic mechanism [4-6]. These models represent acoustic information in a similar way to the auditory nerves and contribute to further understanding of auditory perception. However, implementing such models efficiently in hardware is challenging due to their complex computation and tuning parameters. A higher level of abstraction in the cochlear model, while retaining implementability in low-power electronic technologies, therefore offers a more power-efficient, compact and stable sensor for computing machines and neural implantable devices.

The Spiketrum algorithm, a spike-coding scheme introduced in [7], replicates the human ear's observed mechanism, maintaining relatively low computational complexity. Spiketrum modulates the spike timing and the instantaneous spike rate in a way similar to biological auditory coding and possesses informational robustness to neural fluctuations and spike losses. It provides an efficient coding scheme not only fully compatible with spike-based computational models, but also implementable in hardware for use with neuromorphic hardware technologies. In [8], the hardware implementation details are presented for a real-time, FPGA (Field Programmable Gate Array)-based neuromorphic cochlea, which is founded on the Spiketrum algorithm. It performs real-time processing with a controllable output spike rate, which suits complex machine hearing tasks such as sound classification. The spike rate configuration of Spiketrum hardware is an important factor in determining its performance and usefulness across various user cases.

This paper comprehensively studies the Spiketrum hardware configuration, followed by an evaluation of the performance of both the hardware Spiketrum encoder and its software counterpart against state-of-the-art algorithms such as MATLAB-based Auditory Peripheral (MAP) [9], and to other hardware-efficient algorithms, namely Spectrogram [1] and Lauscher [10]. The paper assesses encoder performance in various sound classification tasks, including classification accuracy, training time, and sparsity. Different classification tasks are demonstrated across various spiking and non-spiking processing neural architectures, namely, Recurrent Spiking Neural Network (RSNN), Long Short-Term Memory (LSTM),

Manuscript received Oct XX, 2026; revised Nov XX, 2026; accepted Dec XX, 2026. Date of publication Jan XX, 2027; date of current version Jan XX, 2027. MHD Anas Alsakkal, Runze Wang, and Jayawan Wijekoon are with the Department of Electrical and Electronic Engineering, The University of Manchester, Manchester, UK. Huajin Tang is with the College of Computer Science and Technology, Zhejiang University, and with the Zhejiang Lab, Hangzhou, China. Corresponding author *Jayawan Wijekoon* (e-mail: jayawan.wijekoon@manchester.ac.uk).

and Convolutional Neural Network (CNN). Further evaluation includes assessing information entropy and gain, hardware resource utilization, and power consumption for the hardware versions of the encoders. The paper provides an application-based analysis of each encoder, comparing trade-offs and using a performance matrix for summarizing remarks.

In addition to the comprehensive evaluation of Spiketrum using kernel sets optimized for audio signals following the methodology in [7], this work also proposes an adaptive kernel learning approach that enables the encoder to adapt its kernel set to signals with different spatiotemporal characteristics. Specifically, the kernel set is updated to minimize the error between the input signal and the signal reconstructed from its spikes, allowing application-specific kernels to be learned while preserving Spiketrum's event-driven encoding mechanism and compatibility with efficient software and hardware implementations. The applicability of this approach is demonstrated using non-audio biomedical signals, namely electrocardiogram (ECG) data, illustrating the generality of the Spiketrum framework beyond audio processing.

## II. Spiketrum and its Hardware Implementation

Spiketrum is an advanced spike-based encoder that transforms continuous signals into neuron-friendly formats. Its architecture comprises three stages: Efficient Encoding, Residual Computing, and Intensity-to-Place Coding (as in Fig. 1-a). In the Efficient Encoding stage, the input signal undergoes matching pursuit signal decomposition using a dictionary of 40 predetermined Gammatone kernels (ϕ(t)), closely mimicking human ear frequency selectivity. Convolution operations identify the best-matching kernels for the captured input audio signal $\hat{x}(t)$ . This frequency-domain process involves computing the Fast Fourier Transform (FFT) of the input audio signal, multiplying it with stored FFT values of the kernel set, and finally, performing an inverse FFT (IFFT) on the multiplied result.

This process generates detailed codes that capture the kernel index (spatial position) m, precise temporal positioning τ, and the correlation intensity between the signal and the matched Gammatone kernel s. Each code (m, τ, s) preserves the distinctive features of the input, and the iterative search encodes the most significant features first. The Residual Computing unit then removes what has been captured by shifting the matched kernel center to τ and scaling it by intensity (s). The shifted scaled kernel is subtracted from the input segment x(t) to calculate the residual $x_{new}(t)$, which overwrites the Signal RAM for a new code generation operation. In the Intensity-to-Place (ITP) Coding stage, codes are mapped onto output channels as binary spike trains, minimizing information loss during the transformation from continuous waveform to discrete spikes. Each kernel is mapped to three output channels with different intensity levels; the circuit selects the channel whose level is closest to the convolution intensity and emits a spike there at temporal position τ. This preserves the spatiotemporal information while a dense continuous signal is translated into a discrete format while retaining the original signal's characteristics [7].

The number of codes generated (i.e., the number of iterations, i) per input audio segment, denoted as sps, is configurable and determines the output spike rate of the Spiketrum hardware across all channels. After completing the specified number of iterations in real-time, the newly captured input audio segment is processed in the same manner. A noteworthy feature of the Spiketrum algorithm is its ability to perform lossless decoding of the generated spikes or codes, facilitating the reconstruction of signals without loss [7]. This capability enables the verification of Spiketrum's ability to preserve temporal and spatial features with minimal information loss. Further, Spiketrum can process diverse inputs, extending its utility to various signal encoding applications beyond audio signals. For more algorithm details, refer to [7].

The Spiketrum hardware is implemented on the XEM7310 Opal Kelly board, equipped with the Xilinx Artix-7 FPGA (Fig. 1-b). The primary hardware consideration is given to real-time operation, striking a balance between accuracy and hardware resources. One key advantage of Spiketrum is its precise control over the output spike rate. This feature enables a fine balance between encoding efficiency and power consumption, making it adaptable to the specific requirements of various applications. For a comprehensive understanding of internal circuits, resources, and optimization techniques within the hardware implementation, please refer to [8].

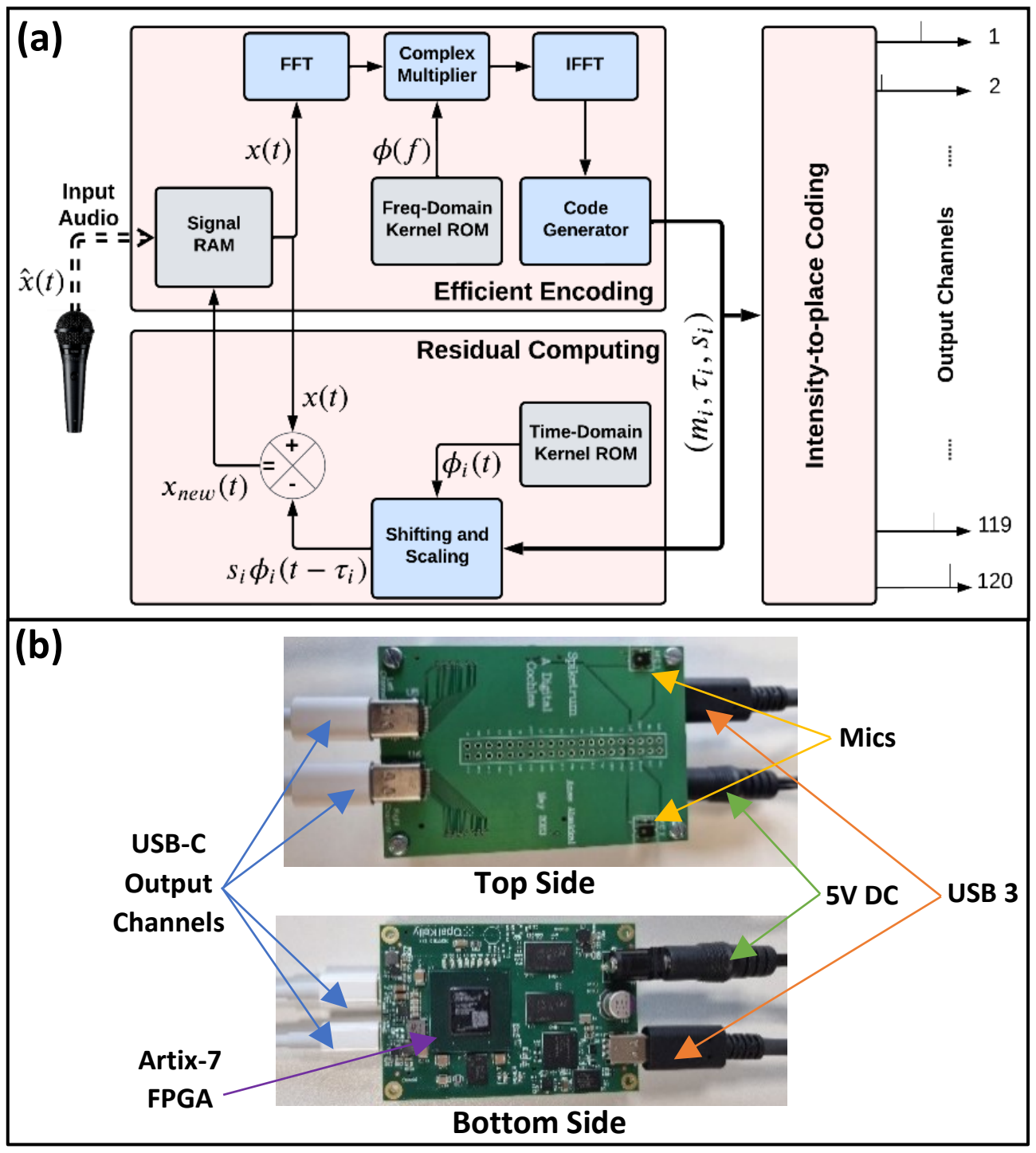

**Fig. 1**. The processing steps of the hardware Spiketrum are presented along with its hardware implementation prototype. **(a)** A block diagram of the processing units of the Spiketrum hardware implementation. **(b)** An FPGA-based Spiketrum hardware. An audio input signal can be streamed into the Signal RAM of the Spiketrum via two microphones in real time, or via a USB-3 interface for offline use with pre-recorded inputs. The 120 spike output channels of the Spiketrum are accessible via a USB-C interface.

On the other hand, the software implementation provides flexibility and ease of use, allowing users to implement Spiketrum on various computing platforms. In this work, the software version is developed in MATLAB using floating-point precision, enabling high numerical accuracy and rapid prototyping.

## III. Methodology

### A. Benchmarked Spike-based Encoders

Spiketrum's performance is assessed by benchmarking the proposed design against other well-known biologically-plausible encoders: MAP [9], Spectrogram [1], and Lauscher [10].

**MAP:** The MAP model [9] implements most of the principles and functionalities found in the human auditory system, encompassing the ear's periphery sensor and neural processing in the brain. The model's ear structure comprises cascading modules emulating outer and middle ear resonances, basilar membrane response, inner hair cell stereocilia coupling, receptor potential, calcium dynamics, transmitter release, and adaptation at the auditory-nerve synapse [11-14].

**Lauscher:** The Lauscher algorithm [10] converts audio waveforms to spikes through three stages. It first achieves spatial frequency dispersion via a simplified basilar membrane (BM) model, extracting key frequency features across 700 channels. Second, intensity is converted to firing rates in hair cells (HCs), producing phase-encoded spike trains. Refractory periods are respected, and bushy cells (BCs) further enhance phase-locking of HCs' output. BCs are leaky integrate-and-fire neurons with feed-forward weights. Details and analysis are available in [10].

**Spectrogram:** The Spectrogram algorithm, as proposed by Dennis et al. [1], extracts local features from the input sound waveforms using time-frequency spectrograms based on 40 Mel filters. Subsequently, a codebook is generated by learning the variations in local feature information through a clustering architecture (such as k-means clustering or self-organized maps). The selected output channel generates spike trains by matching the nearest codebook entry to the extracted local features. Self-organized maps (SOMs) with dimensions of 50x10 neurons are employed for codebook generation. Further insights into the spectrogram algorithm can be found in [1].

### B. Neural Network Classifiers Used

Various classification tasks are demonstrated using non-spiking neural network classifiers (CNN and LSTM), as well as two types of spiking classifiers (RSNNs). Their performances are compared to assess the proposed design against other well-known state-of-the-art encoders.

#### 1) Non-spiking classifiers

This study employed CNN and LSTM, as non-spiking classifiers. These established classifiers serve as baselines for comparison with spike-based classifiers in subsequent sections. These classifiers are built and trained using TensorFlow 1.14.0 [15] and Keras 2.3.0 [16]. The training utilized the ADAM optimizer [17], with an initial learning rate of $10^{-3}$, decreasing by $10^{-4}$ per epoch. Training spanned 100 epochs, with accuracy evaluation each epoch. **CNN:** CNNs excel in image recognition, capitalizing on translation-invariant patterns and spatially-independent hierarchies [18]. The CNN architecture employed here comprises an input layer with P nodes (corresponds to the C output channels of the encoder, which differ between encoders), followed by three 2-D convolutional layers (for flattening and output) with max-pooling, dropout, and Rectified Linear Unit (ReLU) activation. **LSTM:** LSTMs handle long-term dependencies and are chosen for temporal sound analysis. The employed LSTM architecture consists of an input layer with P nodes, a 128-node LSTM layer, a fully connected 128-node ReLU layer, dropout (rate 0.3) for generalization, and a 10-node SoftMax output layer.

#### 2) Spiking classifiers

RSNNs efficiently process temporal information using event-based recurrent connections. The study employs RSNN-based classifiers via two spiking neuron models: Leaky Integrate and Fire (LIF) [10] and Adaptive Leaky Integrate and Fire (aLIF) [19]. The resulting classifiers are denoted RSNN and adaptive RSNN (aRSNN), respectively. The input-to-hidden and hidden-to-output layers are fully connected, and the classifiers were built and trained in SpyTorch [20]. Categorical cross-entropy loss functions are employed, and input spikes are time-binned into 10 ms trains. The 10-node SoftMax-activated output layer produces the 10 distinct audio categories. Training uses supervised learning, Backpropagation Through Time (BPTT), and surrogate gradient descent for loss reduction, enhancing Heaviside function transition in the forward pass.

Table I summarizes the implemented spiking and non-spiking neural network classifiers, detailing their architectures, training configurations, and providing a comparative overview of their performance on Spiketrum-encoded outputs.

TABLE I: A Summary of Classifier Architecture and Training Details

| Classifier | Architecture | Training |
|---|---|---|
| **CNN (non-spiking)** | Input (P), 3× [Conv2D + MaxPool + Dropout + ReLU], Flatten, FC, SoftMax (10) | ADAM, initial LR = 1e-3 with 1e-4/epoch decay, 100 epochs |
| **LSTM (non-spiking)** | Input (P), LSTM (128), FC ReLU (128), Dropout (0.3), SoftMax (10) | As CNN |
| **RSNN (LIF, spiking)** | Fully connected input–hidden–output, LIF neurons, SoftMax (10) | BPTT with surrogate gradients; input spikes binned at 10 ms |
| **aRSNN (aLIF, spiking)** | As RSNN with adaptive-threshold (aLIF) neurons, improving adaptability to varying input dynamics | As RSNN |

Number of input-layer nodes, P = number of encoder output channels, C.

### C. Adaptive Kernel Learning

In addition to employing predefined kernel sets optimized for audio processing, this work extends Spiketrum with an adaptive kernel learning approach that enables adaptation to signals with different spatiotemporal characteristics. The proposed approach adopts a unified encoding–decoding architecture, in which both the spike-based encoder and a linear waveform reconstruction decoder operate over a shared set of kernels $\Phi = \{\phi_1, \dots, \phi_M\}$, where M denotes the total number of kernels in the shared kernel set. Given an input signal $x(t)$, sparse spike events are generated using the Event-based Temporal Matching Pursuit (E-TMP) algorithm [7]. E-TMP represents the signal as a linear superposition of time-shifted kernels selected from $\Phi$,

$$x(t) \approx \sum_{i=1}^{N} s_i \ \phi_{m_i}(t - \tau_i) \quad (1)$$

where $s_i$, $\tau_i$, and $m_i$ denote the amplitude, temporal shift, and kernel index of the $i$-th spike event, respectively, and N denotes the total number of spike events generated for the input signal. This encoding process is greedy and non-differentiable due to the discrete selection of kernel indices and event times.

Signal reconstruction is performed by a differentiable decoder that uses the same kernel set to synthesize the reconstructed waveform,

$$\hat{x}(t) = \sum_{i=1}^{N} s_i \ \phi_{m_i}(t - \tau_i) \quad (2)$$

Kernel learning is therefore driven by minimizing the reconstruction loss,

$$L_{\text{rec}} = \frac{1}{2} \parallel x(t) - \hat{x}(t) \parallel^2 \quad (3)$$

with gradients backpropagated exclusively through the decoder to update the shared kernels $\Phi$. Each kernel is initialized with zero-mean Gaussian noise and normalized to unit L2 norm, then trained with Adam (learning rate 0.001, batch size 64) for 100 iterations under a plain MSE loss with no additional regularization.

Although the encoder is bypassed during gradient computation, the shared-kernel formulation establishes an implicit feedback loop: as the kernels are optimized to reduce reconstruction error, they evolve into more representative temporal basis functions. In subsequent encoding iterations, the E-TMP encoder naturally selects these improved kernels, leading to more efficient and structured sparse representations without requiring direct differentiation through the greedy search process.

This adaptive kernel learning approach enables application-specific kernel sets to be learned directly from data while preserving the event-driven nature of Spiketrum. Importantly, it maintains compatibility with efficient software and hardware implementations, as the encoding algorithm itself remains unchanged.

### *D. Utilized Datasets*

In this study, we present experimental results for sound classification, clustering, and speaker identification tasks, using diverse datasets encompassing various sound types. The utilized datasets, Spoken Heidelberg Digits (SHD) [10], Google Speech Commands V2 (GSC2) [21], Medley DB (MDB) [22], and Forest Sound Classification Dataset (FSC22) [23], offer comprehensive samples for fair benchmarking and thorough encoder evaluation. In addition, the MIT-BIH Arrhythmia Database [24], [25] is used to demonstrate the adaptive learning of the kernel set for ECG signals.

**SHD:** This dataset includes 10,420 spoken digits (0–9) by 12 speakers in English and German. We used 5090 English digit samples, split into training (4011) and testing (1079) subsets. The test subset primarily featured digits from speakers 4 and 5, with a 5% inclusion of trials taken from other speakers, also serving to estimate sparsity and entropy results for all the encoders. **GSC2:** This dataset contains 65,000 samples of 30 English words from diverse individuals from the public. It is used to evaluate encoders' generic feature extraction capability. It includes common words, numerical digits, and directional indicators. We trained the architectures using the SHD dataset and tested on digit samples from GSC2. **MDB:** This dataset offers 3187 pre-annotated sound segments, divided into 14 classes representing musical instruments. A training subset of 1000 samples from these 14 classes was used for the music clustering task. **FSC22:** This dataset contains 2025 labeled 5-second sound samples. It spans six parent-level classes and multiple subclasses, e.g., Mechanical Sounds, Animal Sounds, etc. A uniform collection of 75 audio samples was taken from each of the available 27 subclasses. We extracted 1000 samples primarily from Animal Sounds and Environmental Sounds for sound clustering, combining them with English digit and music instrument sounds obtained from other datasets. **ECG (MIT-BIH):** ECG recordings are sourced from the MIT-BIH Arrhythmia Database, which contains 48 half-hour single-lead ECG recordings sampled at 360 Hz and annotated with multiple heartbeat categories. For the adaptive kernel learning experiments, the ECG signals are segmented into short temporal windows corresponding to individual cardiac cycles, enabling the learned kernels to capture localized physiological morphologies. Ten heartbeat classes are considered, with length approximately 200 samples (556 ms) in each segment. Beats are pooled across all 48 records rather than split on a record-wise basis. For kernel learning, a class-balanced dataset of 500 segments per class (5,000 segments in total) is constructed from the pooled beats. For minority classes with fewer than 500 eligible beats, segments are sampled with replacement to achieve the required class balance.

### *E. Sparsity*

Sparsity is pivotal in spike-based encoding and processing, denoting the percentage of active neurons at a given time. It is important for computational load and energy efficiency: lower sparsity means fewer active neurons and therefore lower energy use. We calculate the average sparsity (using the sparsity definition from [19], Eq. (4)) for each task to assess encoding scheme efficiency—where fewer active neurons convey the same information.

$$Sparsity = \frac{total\ spikes}{total\ neurons\ \times\ time\ steps} \quad (4)$$

### *F. Entropy and Information Gain*

Entropy evaluates information and uncertainty within a system, especially when assessing spike-based encoders. Our study employs an FFT-based entropy method [26], offering versatile measurement of information transmission. It accommodates populations of any size, allowing comprehensive content assessments. To calculate FFT-based entropy, the power spectrum of each neuron's spike train is computed, capturing signal power distribution across different frequencies. Combining these spectra from all neurons forms an overall power spectrum, yielding the determined entropy value. This quantifies information transmitted by the neural population, accounting for correlations among spike trains. Furthermore, we assessed the correlation level between output channels (neurons) in terms of population coding, referred to as information gain.

## IV. Characterization and Benchmarking

This section analyzes the Spiketrum hardware across classification methods and configurations, and compares the selected configuration with other state-of-the-art spike-based encoders in terms of hardware efficiency and effectiveness.

### *A. Spiketrum Hardware Performance Across Different Configurations*

This section explores the impact of Spiketrum encoder output spike rates on hardware properties and assesses its applicability across various use cases. The encoder's spike rate parameter dictates the resolution of the encoded spike output. To guide spike rate selection, this section analyzes Spiketrum hardware outputs for different rates (32, 64, 128, 256, 512, and 1024 spikes per segment (sps) across all channels) with a processed segment size of 43.5 ms. Higher rates capture finer signal features but lengthen processing, raise power consumption and increase resource use, and a higher rate does not by itself guarantee better encoding. Encoding efficiency is assessed through classification effectiveness, entropy and information gain. Most hardware resources remain consistent across configurations, excluding storage units like FIFOs. Power and resource estimates are provided for Spiketrum and other encoders (refer to Section IV-C for an overall comparison). Sparsity and training time are also compared across classifiers.

#### 1) Classification Performance: Accuracy, Training Time and Sparsity

We assess the classification efficiency of Spiketrum's encoded outputs at varying spike rates through a sound classification task, deploying an array of non-spiking (LSTM and CNN) and spiking (RSNN and aRSNN) classifiers. To this end, we employ all samples from the spoken English digits dataset, provided by the SHD dataset [10], encoding them via Spiketrum to generate spikes (and codes for non-spiking classifiers), which are then processed offline by the four implemented classifiers. The 4011/1079 training/test split of Section III-D is used, with sample durations of 0.2343–1.1608 s. The same set of speakers appears in both the training and test partitions, making this a speaker-dependent (person-specific) classification task rather than a generalized speaker-independent evaluation. As anticipated, increasing the spike rate prompts an increase in classification accuracies for all classifiers until the spike rate hits 256 sps (or 512 sps for RSNN), as demonstrated in Fig. 2-a. Notably, non-spiking classifiers outperform their spiking counterparts, although the latter shows an improved tolerance towards oversampled data (>256 sps). This gap arises because CNNs and LSTMs use continuous-valued activations and well-optimized backpropagation, whereas spiking classifiers trade signal detail for efficiency, rely on surrogate gradients that slow convergence, and must additionally encode precise timing.

The optimal accuracy values attained stand at 98%, 99%, 96%, and 95% for CNN, LSTM, RSNN, and aRSNN classifiers, respectively. For context, the chance-level accuracy for this 10-class task is 10%, confirming that all models significantly outperform random guessing, even at lower spike rates. These results were realized at a spike rate of 256 sps for all classifiers, except RSNN which peaked at 512 sps. Fig. 2-b visually presents the number of epochs needed to achieve the maximum accuracy attained per classifier at different spike rates on the test dataset after 100 epochs of training. Intriguingly, classifiers manage to maintain an accuracy exceeding 85% even at the lowest spike rates in all instances, except for RSNN at 32 sps, which remains slightly above 80%. For most cases, CNN demonstrates superior classification accuracy performance over LSTM. Nevertheless, the performance gap narrows with increasing spike rates, and at a rate of 1024, LSTMs noticeably surpass CNNs. Meanwhile, aRSNN significantly outperforms RSNN at lower and moderate spike rates, although its accuracy plunges significantly at higher rates, falling short of RSNN values. The average training duration per epoch stands approximately at 3, 1.28, 28, and 30 seconds for CNN, LSTM, RSNN, and aRSNN, respectively. These times were obtained by averaging the time taken over 60 epochs and are provided for relative comparison, as they depend on the computer's performance.

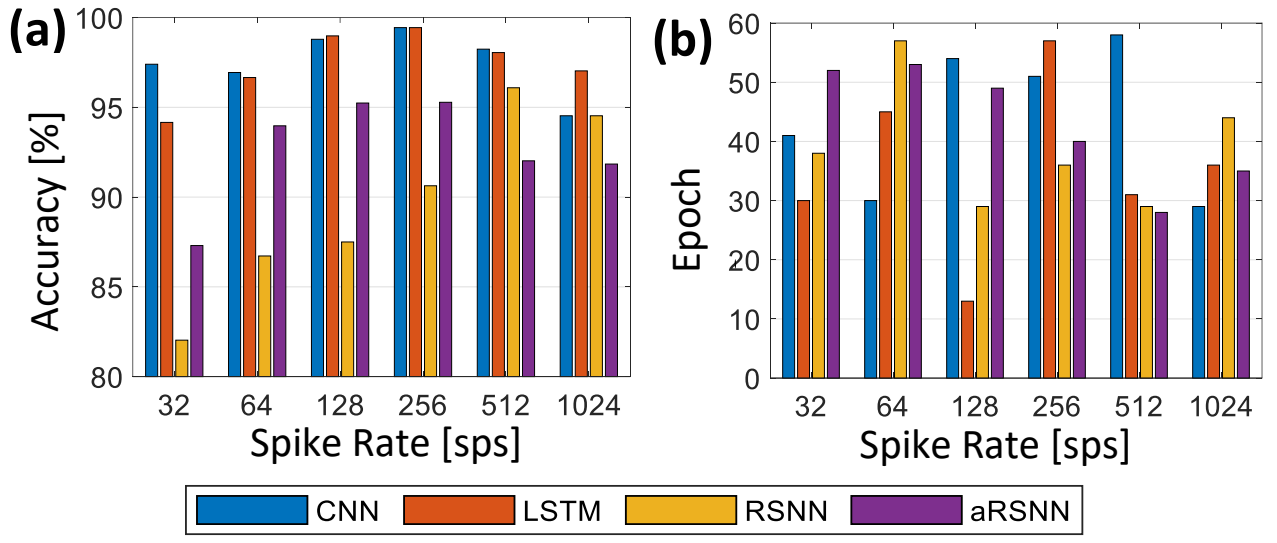

**Fig. 2.** Classification accuracy and training time analysis of the Spiketrum hardware at different spike rates (spikes per segment, sps, is used, and the segment size is 43.5 ms) for spiking and non-spiking classifiers. **(a)** illustrates the maximum classification accuracy on the test dataset after 100 epochs of training for CNN, LSTM, RSNN and aRSNN. **(b)** illustrates the number of epochs required to achieve the maximum accuracy across various spike rates. These results are based on the analysis performed using the SHD dataset [10].

Considering both classification accuracy and training time, a spike rate of 128 sps proves optimal for LSTM and aRSNN. At this rate, LSTM and aRSNN achieve a training accuracy of 85% in just 2 and 9 epochs, respectively. We chose 85% as a reference because most classifiers reach this accuracy exponentially fast before fine-tuning gradually to reach their maximum accuracy. Interestingly, CNN and aRSNN show an inverse proportional relationship between training time and spike rates.

To quantify the sparsity values in our classification tasks using the spiking classifiers (RSNN and aRSNN), we employed the sparsity definition provided in [19] (see Section III for more details). Fig. 3-a illustrates the relationship between sparsity and test accuracy across different spike rates for the spiking classifiers. As shown in Fig. 3-a, the sparsity values of RSNN and aRSNN exhibit variations in response to the spike rate used by Spiketrum. Interestingly, there is a significant disparity between the lowest and highest sparsity values for aRSNN, amounting to approximately a 40% shift. In contrast, the corresponding changes in accuracy are relatively modest, less than 10%. The relationship between spike rate and sparsity is not linear, so the spike rate must be chosen against the accuracy and efficiency requirements of the specific task.

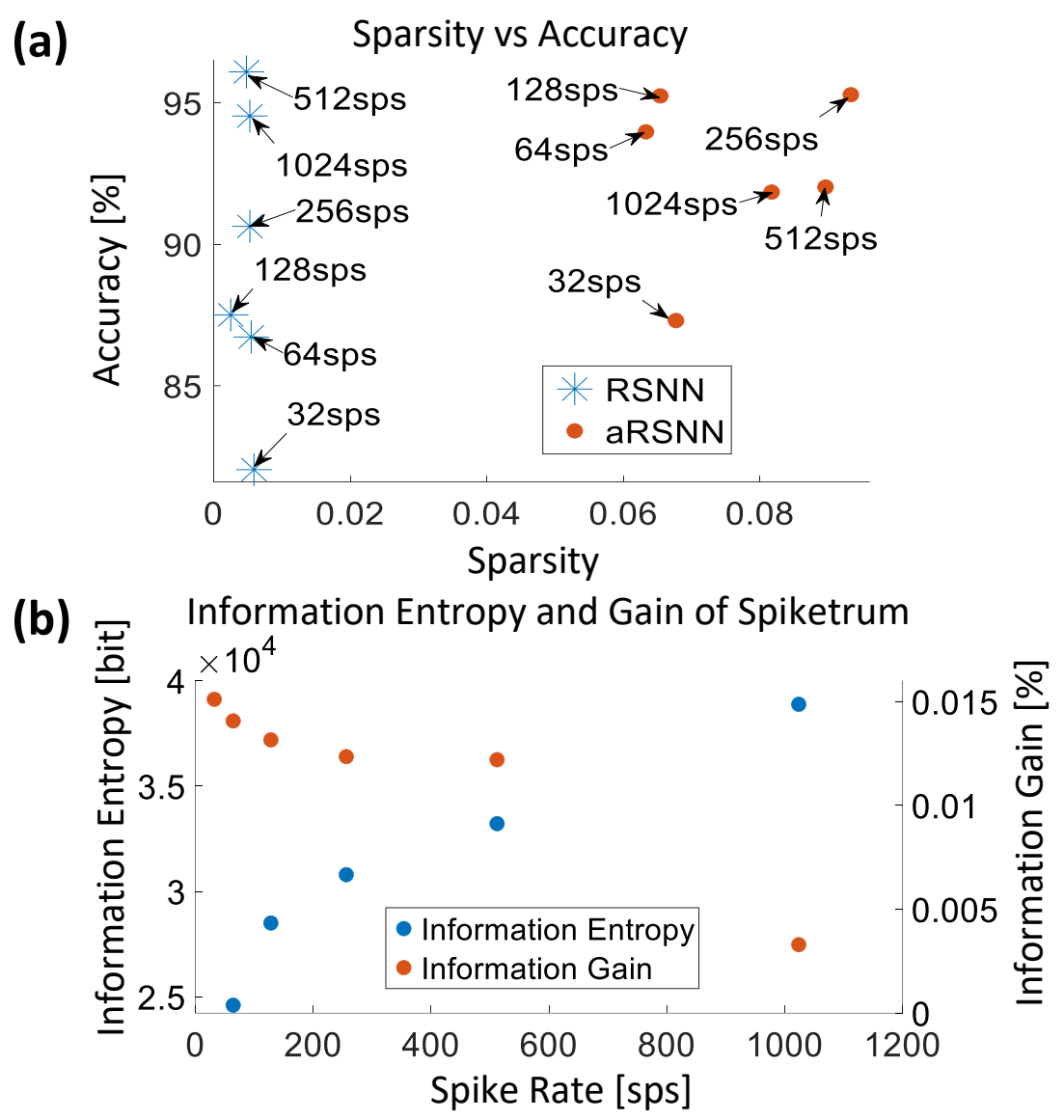

**Fig. 3.** Analysis of Spiketrum Hardware: **(a)** Sparsity vs. test accuracy with two spike-based classifiers. **(b)** Information entropy (left y-axis) and gain (right y-axis) of the encoded spikes vs. output spike rates of the Spiketrum hardware.

### 2) Entropy and Information Gain

Entropy and information gain quantify the information content of the encoded outputs. To estimate entropy, we employ the FFT-based entropy estimation method proposed in [26] (see Section III for more details). Fig. 3-b presents the information entropy and gain of the encoded outputs using the same SHD test subset as the classification task. Increasing the spike rate raises the information conveyed, while the information gain (compared with the other encoders in Section IV-B, Fig. 4-f) remains remarkably low. Information gain represents the correlation level between spike patterns generated by the output channels. The Spiketrum hardware therefore represents information with fewer but more meaningful spikes, minimizing redundant activity.

Table II presents a comprehensive performance summary of various spike rates of Spiketrum over spiking and non-spiking classifiers. The overall performance score is calculated using a weighted scoring method that considers multiple criteria, including classification accuracy, training time, and entropy. Because sparsity only applies to spike-based encoders, we did not consider sparsity in our calculations. Each criterion has been normalized to facilitate a fair comparison across the different configurations. Normalization was performed by scaling each criterion's values linearly to a 0–1 range. Accuracy and entropy are scaled so that the best observed value across all configurations corresponds to 1 and the worst to 0, whereas training time is scaled so that the longest time corresponds to 1, which is why it is subtracted in the overall score.

To determine the overall performance, specific weights have been assigned to each criterion, reflecting their relative importance. Classification accuracy, with a weight of 0.4, captures the configuration's ability to accurately classify input data. The approximate average training time, assigned a weight of 0.3, represents the duration required for the configuration to complete the training process and achieve optimal accuracy, with shorter times being preferred. Additionally, entropy, also assigned a weight of 0.3, serves as a measure of the encoder's information content and efficiency.

The overall score for each encoder configuration is calculated using the following equation:

$$\begin{aligned} Overall\ Score = \\ 0.4\, Accuracy_{norm} + 0.3\, Entropy_{norm} \\ - 0.3\, TrainingTime_{norm} \end{aligned} \tag{5}$$

Where $Accuracy_{norm}$, $Entropy_{norm}$ and $TrainingTime_{norm}$ represent the normalized values of each criterion, scaled to enable fair comparison across metrics with different units and ranges. This holistic score allows for a comprehensive assessment of encoder performance, considering accuracy, training efficiency, and information content. The configuration with 128 sps achieves the highest overall score and is used to benchmark Spiketrum results against other encoders.

TABLE II: Results Summary of Spiketrum Hardware

| Classifier | sps | Accuracy [%] | Epochs | Entropy (×10⁴ bit) | Score |
|---|---|---|---|---|---|
| CNN | 32 | 97.40 | 41 | 2.42 | 0.398 |
| | 128 | 98.79 | 54 | 2.85 | 0.375 |
| | 512 | 98.24 | 58 | 3.32 | 0.395 |
| LSTM | 32 | 94.16 | 30 | 2.42 | 0.442 |
| | **128** | **98.98** | **13** | **2.85** | **0.588** |
| | 512 | 98.05 | 31 | 3.32 | 0.534 |
| RSNN | 32 | 82.03 | 38 | 2.42 | 0.352 |
| | 128 | 87.50 | 29 | 2.85 | 0.459 |
| | 512 | 96.09 | 29 | 3.32 | 0.536 |
| aRSNN | 32 | 87.30 | 52 | 2.42 | 0.300 |
| | 128 | 95.24 | 49 | 2.85 | 0.387 |
| | 512 | 92.02 | 28 | 3.32 | 0.525 |

## B. *Application-Based Evaluation of Spiketrum and Other Spike-Based Encoders*

The comparative analysis of Spiketrum hardware with other state-of-the-art spike-based encoders, including MAP [9], Spectrogram [1], and Lauscher [10], is conducted to assess their performance and resource utilization in applications such as speaker identification, sound classification, and sound clustering. The exercise utilizes the Spiketrum hardware with the 128 sps output spike rate configuration, as it achieves the highest overall performance observed in the previous section. The comparison covers classification accuracy, training time, sparsity, information gain and hardware efficiency, so that the most suitable encoder for a given application can be identified.

### 1) Classification Performance: Accuracy, Training Time and Sparsity

Using several classification tasks and datasets tests how well each encoder extracts and generalizes sound features. The Spiketrum, Spectrogram, Lauscher and MAP encoders were evaluated with all four classifiers.

Fig. 4 presents the classification accuracy, and the corresponding number of training epochs required to achieve the best accuracy for the benchmarked encoders across different implemented classifiers for the four experiments conducted.

The four classification experiments encompassed sound classification based on split cross-validation, sound classification based on external dataset cross-validation, speaker identification, and sound clustering as described below:

#### a) Sound Classification with Split Cross-Validation:

In this task, we conducted a classification experiment with the spoken English digits dataset from the SHD dataset, containing 10 classes. The dataset was divided into training and testing subsets with an 80:20 ratio. We used hardware implementations of the Spiketrum (128 sps), Spectrogram, Lauscher, and MAP encoders to encode the SHD dataset. For the classification task, spiking classifiers utilized generated spikes, while non-spiking classifiers used generated codes. The training process involved the ADAM optimizer [17] as in Section III. Spiking classifiers adjusted the learning rate by 0.5 times the initial rate ($10^{-2}$) every 10 epochs for aRSNN and used a fixed rate of $5 \times 10^{-4}$ for RSNN. Training spanned 100 epochs, with evaluations at every 10 epochs. The CNN and LSTM classifiers were equipped with a dropout layer featuring dropout rates of 0.5 and 0.3, respectively. The same settings were applied to every encoder and classifier. Findings demonstrate strong performance for Spiketrum, Lauscher, and MAP spike encoders across different classifiers, achieving a minimum test classification accuracy of 85% (Fig. 4-a). Spectrogram showed less competitive performance, especially in CNN testing. Among the classifiers used, the LSTM model notably outperformed others, attributed to its exceptional ability to capture long-term dependencies crucial for learning speech patterns. Overall, Spiketrum exhibited very competitive performance across all classifiers, with its hardware implementation reaching peak accuracy in 39 epochs using the LSTM classifier, close to the fastest encoder on this task (Lauscher, 37 epochs). Although training is normally offline, shorter training remains valuable for frequent model updates, transfer learning and faster design iterations.

#### b) Sound Classification with External Dataset Cross-Validation:

To assess the robustness of the extracted features and evaluate the generalization capability of the benchmarked encoders, we conducted an experiment where the four encoders were trained on the GSC2 dataset and then tested on the SHD digit dataset. The training set comprised 6000 samples from 10 classes, representing spoken English digits ranging from 0 to 9.

The SHD test subset was used to evaluate the encoders on this external dataset. As expected in a more challenging experiment, the results demonstrate lower classification accuracy for all encoders compared to the split cross-validation presented earlier (Fig. 4-b), with the exception of Spectrogram, which shows a slight improvement in overall accuracy for both CNN and aRSNN.

This observation could be attributed to the number of features available in the training dataset and the quality of its audio recordings. Among the classifiers, LSTM consistently achieves the best overall results, while RSNN experiences the largest drop in performance. Notably, the highest accuracy achieved, approximately 95%, was observed when LSTM was trained on Spiketrum and MAP encoded spikes. Furthermore, the time taken for the classifier to achieve the best accuracy seems to be similar across all encoders, with no significant differences or outliers in terms of peak training time.

#### c) Speaker Identification:

Speaker identification is a distinct classification task. Audio samples from the 12 speakers in the SHD dataset were used with all classifiers to identify individual speakers. It was split into 3890 samples for the training subset and 1200 samples for the testing subset. To ensure fairness, we maintained an equal distribution of samples per speaker in the training set, with each contributing approximately 300 samples. The results depicted in Fig. 4-c highlight the performance of different encoders in speaker identification. Among them, Spiketrum and MAP both reach the highest accuracy of 99%, with Spiketrum's hardware implementation getting there in fewer epochs (23 with LSTM, against 29 for MAP with RSNN). Spiketrum and MAP therefore produce the most separable spike patterns, and Spiketrum separates speakers better than Lauscher. Spectrogram is the least accurate, although with RSNN it reaches 85% in only 20 epochs.

#### d) Sound Clustering:

This experiment tests whether the encoders can separate broad sound types rather than individual classes, using a combined dataset synthesized from three different datasets: SHD (spoken English digits), MDB (samples of various musical instruments), and Natural Sounds from FSC22 (recordings of natural sounds). The combined dataset, with 3000 samples (2400 for training and 600 for testing), ensures equal representation of each sound cluster. Before encoding, all samples are resampled to 16 kHz.

Results reveal that all encoders achieve accuracy above 80% with relatively short training times (Fig. 4-d), indicating that clustering is an easier task compared to classification. To evaluate classifier spiking activities, sparsity theory from [19] is employed, considering the total spikes and output channels for each encoder. Spiketrum (120 channels), Spectrogram (500), Lauscher (700), and MAP (820) are analyzed. Fig. 4-e compares encoder sparsity with classification accuracy, with Spiketrum's spike rate set to 128 sps. For RSNN, all encoders exhibit low sparsity (0.01 to 0.015), suggesting dense information representation. While Spiketrum hardware reaches the highest accuracy of the three at 89%, Lauscher and Spectrogram follow at 86% and 85%, respectively. Surprisingly, MAP outperforms others significantly at 99% accuracy. For aRSNN, higher sparsity (0.0455 to 0.1145) is observed, indicating a sparser representation. The hardware implementation of Spiketrum (HW Spiketrum) achieves 95% accuracy with approximately 0.0734 sparsity, while the software version (SW Spiketrum) reaches 98% accuracy and 0.0936 sparsity. For comparison, Lauscher at 94% and 0.1145 sparsity, Spectrogram drops to 78% with 0.0455 sparsity, and MAP maintains 98% accuracy with approximately 0.1 sparsity. This suggests that higher sparsity does not necessarily lead to reduced accuracy, with Spiketrum and MAP demonstrating both high accuracy and moderate sparsity. The number of spikes per sample varies significantly across encoders, impacting power consumption and hardware resource requirements. Spiketrum generates far fewer spikes (1,951 per sample) than Lauscher (7,687) and MAP (53,138), which produces the highest count, while Spectrogram generates the fewest (81).

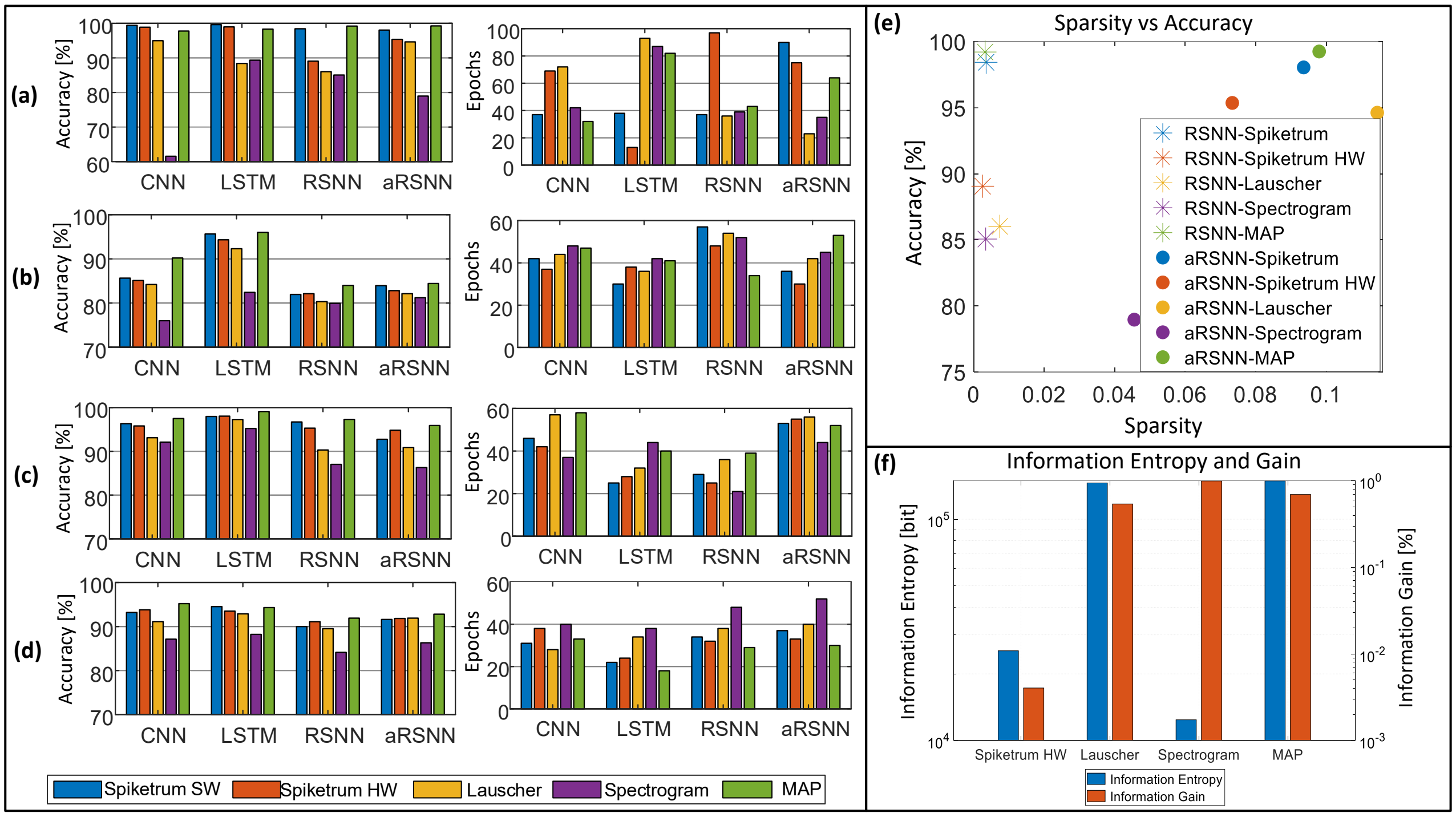

**Fig. 4.** Performance analysis of benchmarked encoders across sound classification and clustering tasks. Left plots display best test accuracy and epochs needed by each encoder for different classifiers to achieve the highest accuracy. Right plots show sparsity and entropy. **(a)** Sound Classification Split Cross-Validation (SHD dataset, 10 classes). **(b)** Sound Classification External Dataset Cross-Validation (GSC2 training, SHD testing). **(c)** Speaker Identification (SHD, 12 speakers). **(d)** Sound Clustering (SHD, MDB, FSC22). **(e)** Sparsity vs. accuracy for RSNN and aRSNN. **(f)** Information entropy and gain (FFT-based) on the SHD test subset for different encoders.

### 2) Entropy and Information Gain

The entropy and information gain of all benchmarked encoders are estimated on the SHD test subset using the FFT-based method of [26]. Fig. 4-f presents the results of the estimated entropy and the information gain. Entropy analysis reveals varying levels of randomness and uncertainty among the benchmarked encoders. Lauscher and MAP exhibit higher entropy, indicating a higher information content encoded. However, Spiketrum demonstrates lower information redundancy, conveying a substantial amount of information with fewer spikes. Notably, the number of output channels plays a significant role, with MAP and Lauscher having a higher number of channels compared to Spiketrum and Spectrogram.

### *C. Hardware Resources and Power Efficiency of Encoders*

Hardware resources and power usage determine whether a spike-based encoder can be deployed in practice, particularly for real-time processing.

To conduct a fair and accurate comparison of the hardware resources and power consumption across various algorithms, a set of assumptions has been made. First, all evaluations are executed using the same hardware and software tools. These include the XEM7310 Opal Kelly board featuring a Xilinx Artix-7 FPGA and Xilinx Vivado 2018.3 software.

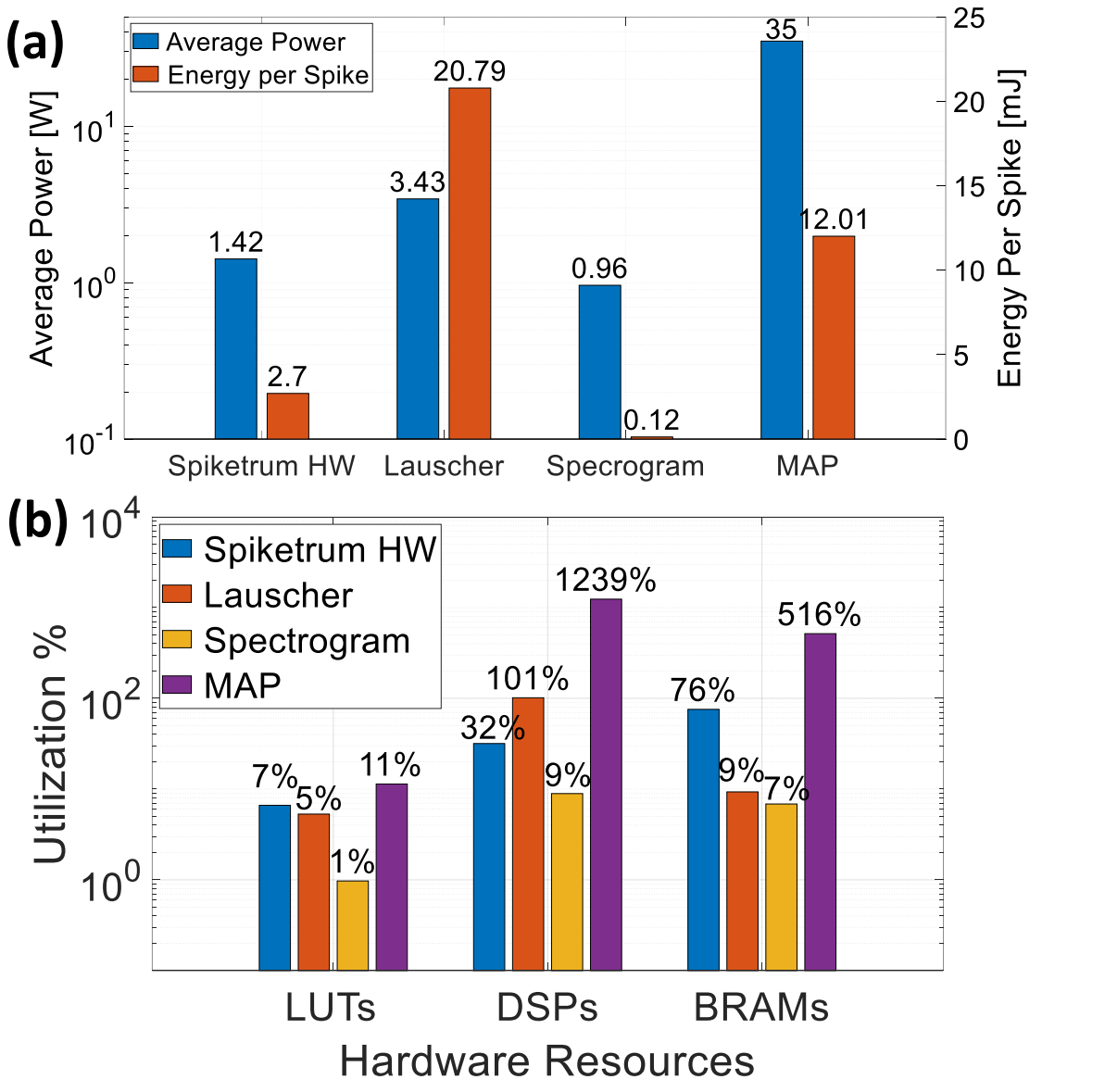

**Fig. 5.** Comparison of Power Consumption and Hardware Resources: **(a)** Average power and energy required to generate a spike for various encoders. **(b)** Hardware utilization estimation on Xilinx Artix-7 FPGA for benchmarked encoders. Power and resource requirements for algorithm processing blocks estimated using Xilinx Vivado 2018.3 tools.

Second, only computations belonging to the encoding scheme are counted, so later neural processing within the cochlea algorithm is excluded. Third, the same design and optimization methods (frequency-based convolution, pipelined architectures) are applied to every implementation. Fig. 5 compares the resulting hardware resources and power consumption.

### *D. Adaptive Kernel Learning for Different Use Cases*

To further evaluate the effectiveness and adaptability of Spiketrum beyond audio processing, the adaptive kernel learning approach is applied to ECG recordings from the MIT-BIH Arrhythmia Database [24], [25] as a representative case study. This experiment assesses whether Spiketrum can learn meaningful temporal basis functions directly from data while preserving sparsity and its event-driven encoding paradigm. A set of $M = 10$ kernels, each 30 samples (83 ms) in length, is learned; the 40-kernel Gammatone set described in Section II serves as the baseline.

As shown in Fig. 6-a, the learned kernels take on structured and interpretable temporal patterns that closely resemble physiologically relevant ECG morphologies. These data-driven kernels enable efficient representation of cardiac dynamics, with distinct and sparse activation patterns observed across different heartbeat types, indicating that the learned kernels capture salient morphological variations.

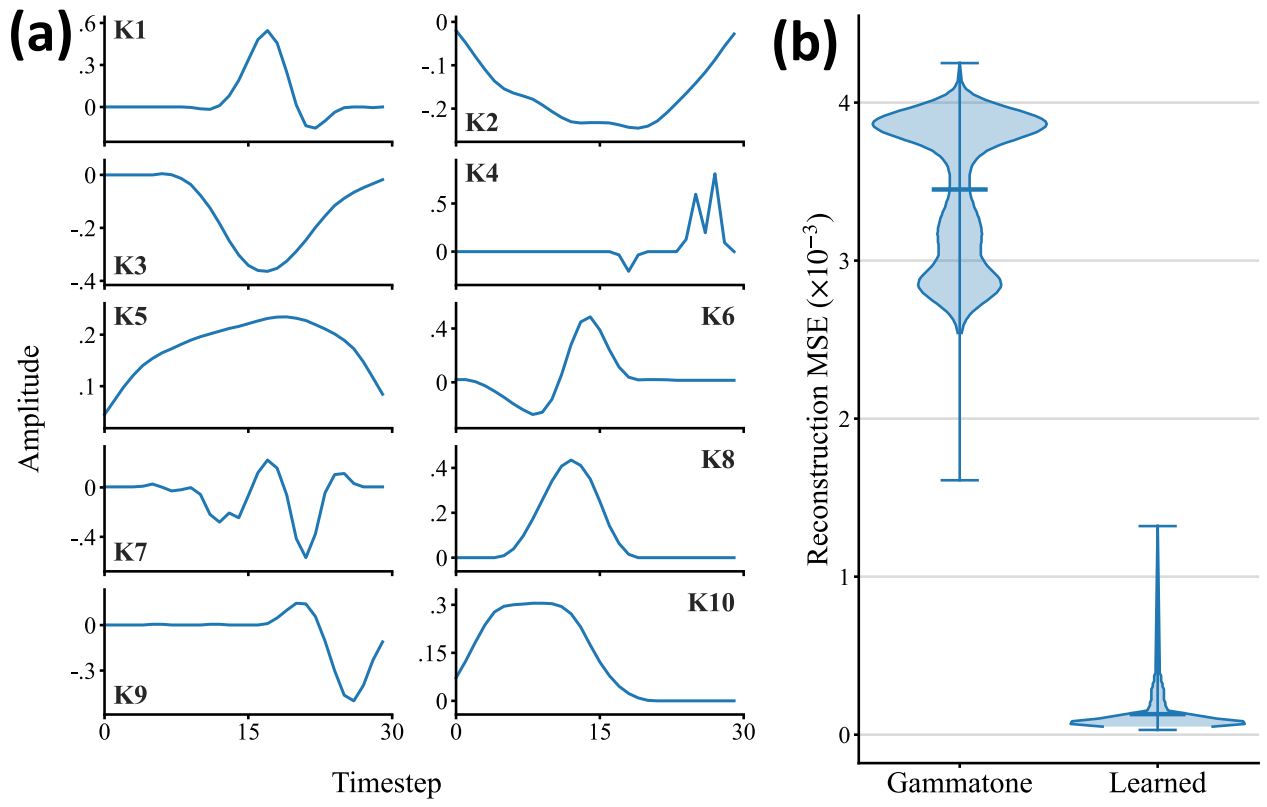

Fig. 6. Kernel optimization results on ECG signals from the MIT-BIH Arrhythmia Database. **(a)** The learned kernel set after adaptive kernel learning. **(b)** Distributions of reconstruction mean squared error (MSE) for Gammatone and learned kernels at a spike rate of 16.

Quantitative reconstruction results are presented in Fig. 6-b. At the same spike rate, the learned kernels yield a lower reconstruction MSE and a narrower error distribution than the Gammatone set, so the signal is represented more accurately without emitting more spikes. The MSE distributions are computed over 10,000 segments randomly sampled from beats not used for kernel learning, without class balancing. Spike rates of 8 and 16 per segment were evaluated. The learned kernels outperformed the Gammatone set at both spike rates; therefore, results are reported at 16 spikes per segment (approximately one spike per 35 ms), corresponding to the higher-quality reconstruction setting.

Overall, Fig. 6 shows that the learned kernels are both physiologically recognizable and measurably more accurate: reconstruction MSE and its variance are lower at an unchanged spike rate. Adaptive kernel learning therefore lets Spiketrum handle signals with fundamentally different spatiotemporal characteristics, such as biomedical waveforms, without altering its core encoding mechanism, extending the framework beyond audio.

## V. Discussion

Section IV analyzes Spiketrum's hardware performance across different configurations and classifiers, with particular emphasis on the spike rate. Table II summarizes the effect of spike rate on output quality, processing efficiency, power consumption, and resource utilization, with 128 sps adopted as the reference configuration. Section IV-B benchmarks Spiketrum, Spectrogram, Lauscher, and MAP, while Section IV-C evaluates their hardware resource requirements and power efficiency, and Section IV-D extends the encoder beyond audio through adaptive kernel learning. The encoders are assessed in terms of classification accuracy, information efficiency through entropy and sparsity, hardware resources, and power consumption, providing a comprehensive basis for practical implementation.

TABLE III: Encoder Performance Across Classifiers and Experiments

| Task | Encoder | Accuracy [%] | Best classifier | Epochs |
|---|---|---|---|---|
| **Classification (split CV)** | Spiketrum HW | 97 | LSTM | 39 |
| | Lauscher | 96 | LSTM | 37 |
| | Spectrogram | 95 | LSTM | 42 |
| | **MAP** | **99** | **LSTM** | **40** |
| **Classification (external CV)** | Spiketrum HW | 94 | CNN | 42 |
| | **Lauscher** | **93** | **LSTM** | **32** |
| | Spectrogram | 87 | LSTM | 43 |
| | MAP | 95 | CNN | 58 |
| **Speaker identification** | **Spiketrum HW** | **99** | **LSTM** | **23** |
| | Lauscher | 95 | LSTM | 33 |
| | Spectrogram | 89 | LSTM | 38 |
| | MAP | 99 | RSNN | 29 |
| **Sound Clustering** | **Spiketrum HW** | **94** | **LSTM** | **12** |
| | Lauscher | 92 | LSTM | 91 |
| | Spectrogram | 82 | LSTM | 86 |
| | MAP | 96 | LSTM | 82 |

Table III summarizes the benchmarking results in terms of accuracy, best-performing classifier, and training time, with faster architectures preferred when accuracy is comparable. CNN and LSTM generally achieve the highest classification accuracies, while RSNN and aRSNN show slightly lower performance. LSTM performs particularly well for speaker identification and sound clustering; however, its reduced generalization to external datasets highlights the need for further optimization. Across the encoders, MAP and Spiketrum achieve strong overall performance, with Spiketrum combining competitive accuracy with a low average spike count and consequently reduced computational and energy requirements.

Entropy analysis shows that Lauscher and MAP produce higher entropy than Spiketrum and Spectrogram, indicating greater encoded information. However, Spiketrum achieves comparable classification performance with lower information gain, suggesting reduced redundancy and more efficient information representation. Together with its sparsity, this enables a lower computational and power burden while retaining the information required for classification.

From a hardware perspective, Spiketrum achieves competitive accuracy with substantially fewer resources than Lauscher and MAP, which require significantly greater hardware resources for comparable performance. Spectrogram has the lowest resource requirements but also consistently the lowest accuracy. Spiketrum therefore provides a favorable balance between accuracy, resource utilization, and power consumption. Its ability to extract both frequency and spatiotemporal features while generating a sparse and controllable spike train further contributes to its efficiency, supporting its suitability for practical spike-based encoding systems. Finally, Section IV-D shows that this efficiency is not specific to audio: a learned ten-kernel set lowers the ECG reconstruction error relative to the forty-kernel Gammatone dictionary at the same spike rate, so the encoder can be retargeted to a new signal domain by reconfiguring its kernels rather than redesigning the hardware.

## VI. Conclusion

This paper shows that Spiketrum offers an efficient, versatile encoding front end for sound-based AI. Across four classification and clustering tasks, it achieved 94–99% accuracy, matching or coming within two percentage points of MAP and outperforming Lauscher and Spectrogram in every task. It combines this performance with sparse output: in sound clustering, it generated 75% fewer spikes per sample than Lauscher and 96% fewer than MAP, while using around 83% and 85% fewer output channels, respectively. Its configurable spike rate provides a direct way to balance accuracy and encoding cost, with 128 spikes per segment giving the best overall balance in this study. Compatibility with both spiking and conventional classifiers makes the same representation useful across different AI systems.

These features make Spiketrum a strong choice when data must be captured and encoded efficiently, especially at the edge, where power and hardware resources are limited. Its real-time hardware implementation supports local processing, while its software implementation enables flexible analysis and integration with broader computing workflows. Beyond audio, adaptive kernel learning reduced ECG reconstruction error at the same spike rate, showing that the encoder can be retargeted without changing its core mechanism. As AI increasingly spans edge devices and cloud platforms, Spiketrum offers a practical way to reduce the cost of turning raw signals into useful machine-readable data.

## References

[1] J. Dennis, H. D. Tran, and H. Li, "Combining robust spike coding with spiking neural networks for sound event classification," in *Proc. IEEE Int. Conf. Acoust., Speech Signal Process. (ICASSP)*, 2015, pp. 176–180.

[2] R. Gütig and H. Sompolinsky, "Time-warp-invariant neuronal processing," *PLoS Biol.,* vol. 7, no. 7, p. e1000141, 2009, doi: 10.1371/journal.pbio.1000141.

[3] P. B. Schafer and D. Z. Jin, "Noise-robust speech recognition through auditory feature detection and spike sequence decoding," *Neural Comput.,* vol. 26, no. 3, pp. 523–556, 2014, doi: 10.1162/NECO_a_00557.

[4] M. Holmberg, *Speech Encoding in the Human Auditory Periphery: Modeling and Quantitative Assessment by Means of Automatic Speech Recognition*. VDI Verlag, 2009.

[5] M. S. Zilany, I. C. Bruce, and L. H. Carney, "Updated parameters and expanded simulation options for a model of the auditory periphery," *J. Acoust. Soc. Am.,* vol. 135, no. 1, pp. 283–286, Jan. 2014, doi: 10.1121/1.4837815.

[6] M. Rudnicki, O. Schoppe, M. Isik, F. Volk, and W. Hemmert, "Modeling auditory coding: From sound to spikes," *Cell Tissue Res.,* vol. 361, no. 1, pp. 159–175, Jul. 2015, doi: 10.1007/s00441-015-2202-z.

[7] H. Tang *et al.*, "Neuromorphic auditory perception by neural spiketrum," *IEEE Trans. Emerg. Topics Comput. Intell.,* 2024.

[8] M. A. Alsakkal and J. Wijekoon, "Spiketrum: An FPGA-based implementation of a neuromorphic cochlea," *IEEE Trans. Circuits Syst. I, Reg. Papers,* vol. 72, no. 4, Apr. 2025, doi: 10.1109/TCSI.2025.3526585.

[9] R. Meddis *et al.*, "A computer model of the auditory periphery and its application to the study of hearing," in *Basic Aspects of Hearing*. New York, NY, USA: Springer, 2013, pp. 11–20.

[10] B. Cramer, Y. Stradmann, J. Schemmel, and F. Zenke, "The Heidelberg spiking data sets for the systematic evaluation of spiking neural networks," *IEEE Trans. Neural Netw. Learn. Syst.,* vol. 33, no. 7, pp. 2744–2757, Jul. 2022, doi: 10.1109/TNNLS.2020.3044364.

[11] E. A. Lopez-Poveda and R. Meddis, "A human nonlinear cochlear filterbank," *J. Acoust. Soc. Am.,* vol. 110, no. 6, pp. 3107–3118, Dec. 2001, doi: 10.1121/1.1416197.

[12] R. Meddis, "Auditory-nerve first-spike latency and auditory absolute threshold: A computer model," *J. Acoust. Soc. Am.,* vol. 119, no. 1, pp. 406–417, 2006, doi: 10.1121/1.2139628.

[13] R. Meddis, L. P. O'Mard, and E. A. Lopez-Poveda, "A computational algorithm for computing nonlinear auditory frequency selectivity," *J. Acoust. Soc. Am.,* vol. 109, no. 6, pp. 2852–2861, Jun. 2001, doi: 10.1121/1.1370357.

[14] C. J. Sumner, E. A. Lopez-Poveda, L. P. O'Mard, and R. Meddis, "A revised model of the inner-hair cell and auditory-nerve complex," *J. Acoust. Soc. Am.,* vol. 111, no. 5, pp. 2178–2188, May 2002, doi: 10.1121/1.1453451.

[15] M. Abadi *et al.*, "TensorFlow: Large-scale machine learning on heterogeneous distributed systems," *arXiv:1603.04467*, 2016.

[16] F. Chollet *et al.*, "Keras," 2015. [Online]. Available: https://keras.io

[17] D. P. Kingma and J. Ba, "Adam: A method for stochastic optimization," *arXiv:1412.6980*, 2014.

[18] Y. Su, K. Zhang, J. Wang, and K. Madani, "Environment sound classification using a two-stream CNN based on decision-level fusion," *Sensors,* vol. 19, no. 7, p. 1733, Apr. 2019, doi: 10.3390/s19071733.

[19] B. Yin, F. Corradi, and S. M. Bohté, "Effective and efficient computation with multiple-timescale spiking recurrent neural networks," in *Proc. Int. Conf. Neuromorphic Syst.*, 2020, pp. 1–8.

[20] F. Zenke, "SpyTorch," 2019. [Online]. Available: https://github.com/fzenke/spytorch

[21] P. Warden, "Speech commands: A dataset for limited-vocabulary speech recognition," *arXiv:1804.03209*, 2018.

[22] R. M. Bittner, J. Salamon, M. Tierney, M. Mauch, C. Cannam, and J. P. Bello, "MedleyDB: A multitrack dataset for annotation-intensive MIR research," in *Proc. Int. Soc. Music Inf. Retrieval Conf. (ISMIR)*, 2014, pp. 155–160.

[23] M. Bandara, R. Jayasundara, I. Ariyarathne, D. Meedeniya, and C. Perera, "Forest sound classification dataset: FSC22," *Sensors,* vol. 23, no. 4, p. 2032, Feb. 2023, doi: 10.3390/s23042032.

[24] G. B. Moody and R. G. Mark, "The impact of the MIT-BIH arrhythmia database," *IEEE Eng. Med. Biol. Mag.,* vol. 20, no. 3, pp. 45–50, 2001, doi: 10.1109/51.932724.

[25] A. L. Goldberger *et al.*, "PhysioBank, PhysioToolkit, and PhysioNet: Components of a new research resource for complex physiologic signals," *Circulation,* vol. 101, no. 23, pp. e215–e220, 2000, doi: 10.1161/01.CIR.101.23.e215.

[26] M. Crumiller, B. Knight, and E. Kaplan, "The measurement of information transmitted by a neural population: Promises and challenges," *Entropy,* vol. 15, no. 9, pp. 3507–3527, 2013, doi: 10.3390/e15093507.